\documentclass[12pt,preprint]{aastex}

\begin{document}

\title{Luminosity Functions of Young Clusters: Modeling the Substellar Mass Regime}

\author{P.R.\ Allen, D.E.\ Trilling}
\affil{\small University of Pennsylvania, 209 South 33rd Street, Philadelphia, PA 19104; pallen@hep.upenn.edu,trilling@astro.upenn.edu}
\author{D.W.\ Koerner}
\affil{\small Northern Arizona University, Dept.\ of Physics and Astronomy, PO Box 6010, Flagstaff, AZ 86011-6010; koerner@physics.nau.edu}
\author{I.N.\ Reid}
\affil{\small Space Telescope Science Institute, 3700 San Martin Drive, Baltimore, MD 21218; inr@stsci.edu}

\begin{abstract}

We predict near-infrared luminosity functions of young (5 Myr to 1 Gyr) star clusters by combining evolutionary models of very low-mass ($1~M_J$ to $0.15~M_{\odot}$) dwarfs with empirical bolometric corrections.  We identify several characteristic features in our results.  These can be attributed to three causes: (1) deuterium burning in the most massive substellar objects; (2) methane absorption in bodies with $T_{eff}$ less than 1300~K, the temperature of the L/T transition; and (3) the formation of dust clouds and the rainout of dust at roughly the same effective temperature as methane formation.  Accurate reconstruction of the substellar mass function from luminosity function observations requires that these phenomena are taken into account. At present, few observational studies extend to sufficient sensitivities to allow detection of these effects. However, the luminosity function of the young open cluster IC~2391 shows a clear peak at $M_I \sim 14$ which we attribute to the result of deuterium burning in substellar objects. The location of this feature is a strong function of age, and we estimate an age of 35~Myr for IC~2391. This is significantly younger than the 53~Myr derived from the location of the lithium depletion boundary but agrees with the main sequence turnoff age. We consider the implications of this result and our multi-band luminosity functions for future observational studies. All predicted luminosity function features are, or will be, accessible to observations using new wide-field IR imagers and the Space Infrared Telescope Facility.

\end{abstract}

\section{Introduction}

The determination of the stellar Initial Mass Function (IMF) has been a prominent goal of astronomy for over 50 years.  The IMF, ${\Psi}(M)$, describes the number of stars born per unit mass per unit volume.  The pioneering study by \citet{salp} was limited to stars with masses greater than $0.3~M_{\odot}$.  Salpeter found that the results were well represented by a power law, $\Psi(M) \propto M^{-\alpha}$, with $\alpha = 2.35$.  Two decades later, \citet{ms79} used improved observations of lower luminosity stars to show that the IMF deviates from a simple power law below $1~M_{\odot}$.  They derived a better fit with a log-normal distribution, $\Psi(M) \propto \exp{(\frac{\log(M) - \log(M_0)}{\sqrt{2}\sigma})^{2}}$, with $M_0 \sim 0.15~M_\odot$, $\sigma \sim 0.6$, and a distribution close to the Salpeter power-law at masses exceeding $1~M_{\odot}$.  The change in slope at lower masses is evident both in field-star surveys and in observations of young open clusters (e.g., \citet{scalo98}).  More recent studies \citep{rg97,kru,inr03} favor modeling the field-star IMF using multiple power-laws, with $\alpha \sim 2.3$ at $M> 1~M_\odot$ and $\alpha \sim 1$ at lower masses.  

Extending coverage to masses close to and below the hydrogen-burning limit has only become possible within the last decade, as improvements in detector and computational technologies have permitted the first high-sensitivity, wide-field near-infrared sky surveys. Analysis of those surveys has led to the discovery of a host of very low-mass ultracool dwarfs  \citep{kp00,burg02,geb02} and the first estimate of the brown dwarf mass function in the field \citep{inr99}.

There are a number of complications in deriving $\Psi(M)$ for substellar mass objects.  Mass is generally not observed directly, but is estimated from the measured luminosity.  In the case of main-sequence stars this calculation is relatively straightforward,  since hydrogen fusion leads to a well defined mass-luminosity relation. Brown  dwarfs, however, lack a stable, long-term energy source, and, as a result, evolve rapidly to lower temperatures and lower luminosities. Thus, the substellar luminosity function depends strongly on both mass and age.  Moreover, the presence of broad spectral absorption features stemming from the onset of molecule and dust cloud formation can lead to significant changes in broadband absolute magnitudes over small ranges of effective temperature. Finally, short periods of deuterium fusion occur in objects with masses greater than $13~M_J$, leading to further transient effects in the luminosity function.

We invert the problem. Starting with an assumed mass function and age distribution, we combine theoretical predictions of the evolution of very low-mass dwarfs from the models computed by \citet{bur} with empirical estimates of bolometric corrections from \citet{dahn} and \citet{rc} to predict luminosity functions for young (5 Myr to 1 Gyr) clusters.  Young clusters are convenient for mass function studies because they offer a large range of masses with similar ages and distances.  \citet{mll} used the \citet{bcah98} models to perform a similar analysis, to masses as low as 0.02 M$_{\odot}$, and probed a complementary age range ($<$~10~Myr) to the current work.  

The outline of this paper is as follows: $\S$2 provides a summary of the input evolutionary models and bolometric corrections for our analysis; $\S$3 describes the main features of the predicted luminosity functions. In $\S$4, we compare those results against observations of several young clusters, and consider the potential offered by future observations.

\section{Modeling the Substellar Luminosity Function}

\subsection{Substellar Evolutionary Models}

Most current studies of very low-mass dwarfs are based on one of two sets of theoretical models.  \citet{bcah98} combine interior models with detailed model atmosphere calculations, but cover only a limited range of mass and effective temperature (essentially corresponding to a lower spectral type limit of L); \citet{cbah00} update these models to cover a lower range of masses, 0.1 M$_{\odot}$ to 0.01 M$_{\odot}$ (spectral types M, L, and T).  In contrast, \citet{bur} present a more extensive series of models, spanning masses between $1~M_J$ and $0.15~M_{\odot}$ ($1~M_{\odot} = 1047~M_J$) and ages in the range $10^6$ to $10^{10}$ years, that do not include model atmospheres, but predict bolometric luminosities and effective temperatures as a function of time. Since our aim is to include coverage of very low-mass stars through low-temperature brown dwarfs, we base our analysis on the \citet{bur} models. 

Figure 1 shows the time variation in effective temperature predicted by the Burrows models. These models provide a series of reference points for specific masses at particular ages.  To achieve the numerical resolution needed for our study, we interpolate among the predicted values of bolometric luminosity and effective temperature using a bicubic spline interpolation as a function of mass and age.  This interpolation steps over fine increments in both mass (${\Delta}m = 1{\times}10^{-4}M_{\odot}$) and age (${\rm \Delta\log (Age(Gyr)) = 8.602{\times}10^{-4}}$).  We do not attempt to model clusters with ages of less than 2--3 Myr, since the model calculations are unreliable at these ages \citep{bur}.

Our goal is to compare the model predictions against observations of near- and mid-infrared ($IJHKM$) luminosity functions of young stellar clusters.  As a first step, bolometric luminosity functions are generated. This process requires assumptions concerning both the IMF and the distribution in age.  We parameterize the IMF as a two-segment power-law, ($\alpha_1, \alpha_2$).  The higher-mass segment includes the range $0.08~M_{\odot}$--$0.15~M_{\odot}$, with $\alpha_1 = 1.05$ held fixed (following \citet{rg97}).  At lower masses, from $1~M_J$ to $0.08~M_{\odot}$, $\alpha_2$ is allowed to vary as a free parameter.  The two segments of the power law are connected at the joining mass, $m_{12}$.  For the set of models presented here $m_{12}$ is fixed at $0.08~M_{\odot}$ (see $\S$4.1).

The age distribution assumption is simplified because a single age is assigned to a cluster.  However, star formation is not an instantaneous event, but rather a continuous process that occurs over a finite time interval. We assume that the ages for stars/brown dwarfs in each cluster uniformly span a range corresponding to $\pm 10$\% of the assigned cluster age.  Although this assumption produces an unrealistically large age dispersion for older clusters, the luminosity variations over the 10\% age interval should not be large enough to affect our predictions.

Combining the IMF and the assumed age range gives a continuous distribution of objects as a function of mass and age.  Each model mass--age distribution is normalized such that the space density of 0.1~M$_{\odot}$ objects is equal to that derived in \citet{rg97}.  This arbitrary normalization is used throughout this work unless otherwise stated.  Lastly, we use the \citet{bur} models to assign values for the appropriate bolometric luminosity and effective temperature at each point in the mass--age distribution.  The luminosity function is derived from this distribution by summing the relative number of objects in each interval of luminosity.

\subsection{Bolometric Corrections}

In order to compare model luminosity functions to observed data, our derived bolometric luminosity functions must be converted into broad-band luminosity functions.  This conversion requires the use of bolometric corrections (BCs).  Empirically derived BCs are more reliable than colors generated by model atmospheres, e.g., \citet{bcah98} (for L and M dwarfs) or \citet{bur}.  Model BCs are only as good as the current knowledge of line opacities of the atoms, ions, and molecules present in brown dwarf atmospheres, whereas empirical BCs have no such model dependencies.  

There are two sets of empirical near-infrared very low-mass star and substellar BCs available \citep{leg02,dahn}.  The temperature scales derived by these two groups differ slightly, with the Dahn effective temperatures closer to those for the Burrows models.  We therefore use the $J$-band BCs from \citet{dahn} and derive $IHK$-band BCs from the \citet{dahn} $IJHK$ colors.  However, the \citet{dahn} BCs lack points in the T dwarf regime, T $<$ 1300K, so we include the data on Gl 229B and Gl 570D from \citet{leg02} in our analysis.  For the $M$-band we use BCs from Table 2 of \citet{rc}, plus spectrophotometric data on Gl 229B from \citet{noll}. 

A combination of polynomials and line segments was used to create a fit of BC as a function of effective temperature in each bandpass (Figure~2).  T~dwarf data (the coolest one $M$-band and two $K$-band points) are joined by straight line segments to the rest of the fit to accommodate the sharpness of the L/T transition.  The effect at $K$-band of the rapid onset of methane formation is seen at the L/T transition temperature of 1300~K (Figure~2, top panel).  This effect leads to a sharp reduction in the $K$-band luminosity over a relatively small range in effective temperature, with a correspondingly significant effect on the $K$-band luminosity function, as is discussed in more detail in $\S$3.2. Below the last data points available we extrapolate a flat line segment.  This reflects our ignorance of the BCs below $\sim$900~K.  However, it is possible that additional structure exists at faint magnitudes in the luminosity function.  We use our derived $IJHKM$ BC fits to derive broadband magnitudes for each distribution of objects in our simulations and construct the $IJHKM$-band luminosity functions in the same manner as the bolometric luminosity functions.

\section{Model Results: Cluster Luminosity Function Morphology}

\subsection{The Bolometric Luminosity Function}

Through application of the methods described in $\S$2, we construct a library of model luminosity functions based on objects in the \citet{bur} mass range with ages between 5 Myr and 1 Gyr.  We show in Figure~3 an example of the general results: the evolution of the bolometric luminosity function of a cluster from 10~Myr to 1~Gyr for a model with power law index $\alpha_2 = 1.0$.  We identify in these luminosity functions significant features that are discussed below.

At 10~Myr, the  cluster exhibits a strong peak (A) at bright magnitudes, a broad trough (B) centered on $M_{bol} \sim 14$, and a broader peak (C) at fainter magnitudes.  Peak A is the low-mass tail of the main sequence ($M > 0.075~M_{\odot}$) together with brown dwarfs sufficiently massive to burn deuterium ($13~M_J < M < 0.075~M_{\odot}$).  Note that the sharp cutoff at bright magnitudes is due to the high mass limit of the Burrows models, $0.15~M_{\odot}$, and is therefore unphysical. The presence of deuterium fusion in higher-mass brown dwarfs maintains their surface temperatures and luminosities at relatively high levels (see Figure 1). In contrast, the non-deuterium burning brown dwarfs evolve rapidly to lower temperatures and luminosities.  This causes the two populations to spread apart, forming a trough (B). Peak~C represents the lowest mass objects ($M < 13~M_J$), which cool continuously after their formation.

The 100 Myr cluster luminosity function has a feature which is not present in the 10 Myr luminosity function: a sharp spike (D) between peaks A and C. By 100 Myr, nearly all brown dwarf deuterium burning has ceased, and these higher-mass brown dwarfs join their lower-mass cousins in rapid cooling. This leads to the formation of the second peak/trough feature (D and E) as brown dwarfs separate from hydrogen-burning stars.  To verify that peak D objects correspond to higher-mass brown dwarfs, we segregate the 100~Myr sample by mass. Figure~4 confirms that peak D includes only brown dwarfs that have finished their deuterium burning phase.

Finally, the bottom panel in Figure 3 shows our prediction for a 1 Gyr cluster.  The evolution of peak A has halted, since nearly all stars are stabilized on the main sequence and all brown dwarfs that burned deuterium have separated from this group. Peak~D has broadened since brown dwarfs cool at mass-dependent rates. A new trough (F) lies in the midst of this peak, reflecting the onset of dust cloud formation and dust rainout in the Burrows models (see $\S$3.2).  Note that peak C spans roughly the same magnitude range at each age (all panels of Figure~3). In general, the cooling rates of substellar objects are mass dependent, but the cooling rates of the lowest mass ($< 13~M_J$) brown dwarfs are similar to each other between 5 Myr and 1 Gyr.  This leads to the uniform behavior of peak C members over time.  We do not include dynamical evolution effects in our models.  While these effects are not hugely important for the younger clusters considered here the effects on old clusters, $\sim$1 Gyr, may be important.

Overall, the bolometric cluster luminosity function exhibits complicated time dependent behavior that depends on both the interior and atmospheric evolution of brown dwarfs.  However, the features identified in Figure~3 are dependent mostly on the evolution of the total emergent luminosity (i.e.\ interior evolution).  Therefore, since the interior physics of these models is relatively well understood, we believe that the locations of the major features, peak D and troughs B and E, are defined robustly.  We note that the normalization of the model luminosity functions ultimately is arbitrary and that only the relative heights and locations of these features are important.

\subsection{Broad-band Luminosity Functions}

Broad-band luminosity functions are constructed using the BCs discussed in $\S$2.2.  Figure 5 displays the evolution of $I$- and $K$-band cluster luminosity functions from 10~Myr to 900~Myr\footnote{We can provide luminosity functions using any desired mass function parameters for any bandpass, if we are given or have the appropriate BCs.  Please send requests to {\tt pallen@hep.upenn.edu}.}.  The morphological features discussed in $\S$3.1 and labeled in Figure 3 are also marked in Figure 5.  The gross morphological features of both bands are similar to those of the bolometric luminosity function.  Peak~A lies at the brightest magnitudes modeled here; the sharp cutoff is due to the upper mass limit of the Burrows models.  Peak~D (representing brown dwarfs that burned deuterium) forms and separates from the (hydrogen-burning) main sequence at ages of 10--20~Myr.  Finally, the peak~C objects form a broad distribution at faint magnitudes.

There is one major morphological difference between the luminosity functions at these two bands: there is a large, persistent depression around $M_K$ = 13--15 at all ages (Figure 5b).  This feature is located near trough F in the bottom panel of Figure~3 (the bolometric luminosity function); however, in Figure~5b ($K$-band), this feature is much deeper.  Recall that there is a change in slope for BC$_K$ at $T_{eff}\approx1300$~K (the L/T transition); this slope change is caused by the onset of methane formation (Figure 2).  This causes the brightest T dwarfs to be substantially fainter at $K$-band than the faintest L dwarfs.  The L/T transition occurs over a small range of effective temperatures, leaving few objects in the transition region at any given age and producing the deep trough in the $K$-band luminosity function.  

There is also evidence for a dip in the $I$-band luminosity function around $M_I$ = 18--20 for all ages (Figure 5a).  This feature is due to the presence of dust grains.  At effective temperatures greater than $\sim$1300~K dust exists in a homogeneous mixture in substellar atmospheres and smoothes out many spectral features.  However, below 1300~K -- the temperature also associated with methane formation -- dust rains out of the photosphere or forms clouds (A.\ Burrows 2002, private communication; \citet{tsuji}).  This removes the smoothing effect, allowing the strong atomic and molecular absorption features to dominate.  This results in a rapid decrease in flux throughout the near-infrared over a short range in effective temperature, which in turn creates trough F.  Trough F is most pronounced in the $K$-band, because both rainout and methane formation occur around the same effective temperature.  

Since the effective temperature at which the L/T transition occurs varies as a function of mass and age, the location of the methane feature may also change as a function of mass and age.  The bolometric correction data that we use is based on observations of field brown dwarfs, which are older and therefore more massive than younger objects with same effective temperatures.  Until bolometric corrections can be obtained as a function of time and effective temperature it will be difficult to determine how strongly the position of the trough F depends on mass.

\section{Discussion}

\subsection{Mass Function Parameters}

Our models include one free parameter: $\alpha_2$, the power law exponent at lower masses.  Figure~6 displays $I$- and $M$-band 100~Myr luminosity functions for three different values of $\alpha_2$. As expected, the steeper mass function, corresponding to $\alpha_2$ = 1.2, leads to a larger relative number of low luminosity objects.  In contrast, with $\alpha_2=0.5$, the luminosity function slopes significantly downward from peak A towards fainter objects.  The joining mass, $m_{12}$, can also be varied as a free parameter.  However, varying $m_{12}$ throughout the range of $0.03~M_{\odot}$ to $0.1~M_{\odot}$ produces minimal ($\sim$5\%) changes in the amplitude of the features in the output luminosity functions, so for the models presented here we have fixed $m_{12}$ to $0.08~M_{\odot}$.

\subsection{Comparison with Observational Data}

We test our models against observed cluster data for three young clusters: Upper~Sco \citep{ard00}, IC~2391 \citep{bn99}, and the Pleiades \citep{bouv} (Table 1).  \citet{ard00} obtained $RIZ$ photometry using the 60~cm Schmidt telescope at Cerro Tololo Inter-American Observatory (CTIO) to survey $\sim$14 square degrees of Upper~Sco, finding 138 candidates.  Twenty-two candidates were followed up spectroscopically and 20 were found to be cluster members.  \citet{bn99} used the 4m Blanco and 1.5m telescopes at CTIO to observe $\sim$2.5 square degrees of IC~2391 at $VRI_{c}Z$ to construct their 132 object sample.  They followed up their initial selections by cross referencing with the 2MASS database to obtain more rigorous color cuts.  \citet{bouv} used the Canada-France-Hawaii Telescope (CFHT) to obtain 62 cluster candidates in $\sim$2.5 square degrees of the Pleiades at $R$ and $I$, twenty-six of which are claimed to be very low-mass stars or brown dwarfs.  These 62 candidates were then added to an older sample of $\sim$440 stars \citep{hhj}.   Each group estimated their foreground/background contamination due to photometric selection to be 25\% to 50\%.  Consequently, the data shown here may have significant contamination from non-members, especially at the faint magnitudes, and the conclusions we draw should be viewed as preliminary pending the availability of deeper, less contaminated samples.

Figure 7 displays cluster models, with $\alpha_2 = 0.5$, produced for the specific ages corresponding to each data set.  Complications arise in the comparison of the data to our models because the brightest model magnitude bins are incomplete due to the upper mass limit of the \citet{bur} models.  The spread in age used for each model cluster causes each model magnitude bin to be composed of both massive, old objects and young, less massive objects.  The complication arises at the brightest model bins because the older objects would have masses greater than 0.15$M_{\odot}$, which are not included in the \citet{bur} models.  To account for this, we discard those most luminous model bins (compare Figure~3 with Figure~7).  The brightest reliable model point is then normalized to the corresponding data point.  The resulting match between the model predictions and the data is extremely encouraging, considering that a rigorous fit to the data was not performed.  This gives us the confidence to make further predictions about future cluster observations.  

The youngest of these three clusters, Upper Sco, is at the young edge of our models.  The fit to this cluster is quite good due to the large number of objects detected.  The other two clusters considered, IC~2391 and the Pleiades, each have two different age estimates derived using two independent techniques (see Table~1).  In both cases, the younger age estimate (35~and 100~Myr, respectively) is based on matching the turnoff from the main-sequence against the appropriate stellar isochrones (such as \citet{mey93}). The older age estimate is derived from the measurements of the lithium depletion boundary among low-mass cluster members.  Since all of these dwarfs are fully convective primordial lithium is cycled through the core regions and progressively destroyed at a rate which depends on the central temperature.  Since the latter parameter depends on mass, lithium is depleted more rapidly in higher mass dwarfs. Thus, as a cluster ages, the boundary between dwarfs with and without detectable lithium absorption moves to lower luminosities and later spectral types. Theoretical models can be used to calibrate that variation, and hence estimate cluster ages \citep{staf}.

We can compare the main-sequence fitting and lithium-depletion ages against our estimate based on the predicted morphology of the luminosity function. In the case of the Pleiades, we cannot distinguish between ages of 100 Myr and 120~Myr, since the discriminating features lie 3 to 5 magnitudes fainter than current survey limit. However, the IC~2391 data are clearly a better match to the 35~Myr luminosity function than the 53~Myr function. The most notable feature in the data is the pronounced peak at $M_I \sim 14$ (Figure 7, middle panel).  This corresponds exactly to the position of peak D (high-mass brown dwarfs) in the 35 Myr model. Peak D lies $\sim2$ magnitudes fainter for the 53~Myr age favored by a recent lithium depletion analysis \citep{bn99}.  However, the data displayed for IC~2391 includes both the probable and the possible members, as defined by \citet{bn99}.  The majority of the cluster stars contributing to the peak are only possible members, and further observations are required to determine the extent of background contamination.

To detect features such as peak D, the sample size for an observed cluster must exceed a threshold value dictated by the Poisson uncertainty in each bin.  For IC~2391 this corresponds to $\sim$10--15 stars per bin for a 3$\sigma$ difference.  Therefore, since peak D covers a range of one and a half magnitudes (three model bins), we require there to be a total of 30--45 stars in the range $13.5<M_{I}<15$.  This is on the edge of the current data for IC~2391 with 22 stars in that range.

In general, we note that the location (in magnitude) of peak D is a strong function of age. This may prove a sensitive, albeit model dependent, diagnostic of the age of clusters with ages between 15 Myr and $\sim$200 Myr.  To date, published observations of IC 2391 are the only data sufficiently sensitive to detect peak D.  Further studies, extending to fainter magnitude limits in this and other clusters, are clearly warranted.

\subsection{Prospects for Future Observations}

The sensitivity limits of current young cluster studies, with the exception of IC~2391, preclude observations of peak D and very faint objects.  The limits of the \citet{bouv}, \citet{bn01}, and \citet{ard00} surveys are given by the faintest data points in Figure 7 and correspond to $m_I \sim$ 20--21.  Figure 8 shows $I$-band luminosity functions for clusters with a range of ages and this sensitivity.  The limit of detectability of peak D depends on the distance and age of the observed cluster.  Additionally, peak D only exists in clusters older than 15 Myr.  With current sensitivity limits, relatively few clusters lie close enough to probe the lower end of the luminosity function.  Furthermore, cluster observations are often hampered by extinction (particularly at $I$-band), and determining cluster membership can be problematic.  However, recent advances in near- and mid-infrared instruments, wide-field imagers, and space-based missions offer encouraging prospects to circumvent these difficulties.

Infrared observations of very low-mass stars and brown dwarfs are advantageous because these objects are quite red in optical$-$infrared colors ($I-JHKM$), and incompleteness effects due to extinction are reduced.  Additionally, at near-infrared wavelengths very low-mass stars and brown dwarfs are close to their blackbody peaks. Of the three primary near-infrared bands ($JHK$), $J$-band is best suited for these studies (Figure~9).  This is because $J$-band lacks the severe methane absorption feature which causes great distortions in the $K$-band luminosity function around the L/T transition and depresses $K$-band fluxes \citep{naka}.  $H$-band is intermediate between $J$ and $K$, and while the methane absorption is not as severe as that in $K$-band, there is some some methane absorption at $H$-band wavelengths.  The combination of these reasons make the $J$-band the ideal choice for future cluster observations from the ground.

Several ground-based near-infrared imagers that have recently come online provide sufficient sky coverage and sensitivity that significant portions of nearby clusters can be mapped efficiently.  For example, Flamingos on Gemini South has a FOV = $2.6'{\times}2.6'$ and reaches $m_J=20.3$~in~300~sec (5$\sigma$~point source detection) \citep{re}.  WIRC-2K (Wide-field InfraRed Camera) on Palomar has a FOV = $9'{\times}9'$ and reaches $m_J = 19.9$ in 300~sec (5$\sigma$ point source detection) \citep{wils}.  With these and other current and forthcoming instruments it will be possible to probe 3 to 5 magnitudes deeper than past surveys, detect peak~D, and better determine the behavior of the very low-mass/substellar mass function.

Although mid-infrared observations yield the best possibilities for detection of peak D and troughs B and E (Figure 10), ground-based mid-infrared observations are difficult because atmospheric water bands limit sensitivity. The forthcoming Space Infrared Telescope Facility (SIRTF) mission will probe to sufficient sensitivities to detect these intermediate predicted cluster luminosity function features.  Figure~10 shows our $M$-band ($4.8~\mu$m) models with the SIRTF InfraRed Array Camera (IRAC) Channel 2 ($4.5~\mu$m) sensitivity limits ($M_M = 13$) for a cluster at 200 pc.  The predicted sensitivity limit corresponds to a 5$\sigma$ detection of a $m_M=19$ point source in 200~seconds \citep{hora}.  For this sensitivity and distance both troughs B and E will be easily detected by IRAC/SIRTF, as will be peaks D and C for clusters younger than $\sim$200 Myr.

\section{Summary}

We have combined theoretical evolutionary tracks for low-mass dwarfs with empirical bolometric correction measurements to identify several characteristic features in the predicted luminosity functions at $IJHKM$-bands.  The most significant feature that we find is an intermediate peak in clusters that are tens to hundreds of millions of years old (denoted as peak D) corresponding to high-mass brown dwarfs that have ceased to burn deuterium. This feature appears to be present in observations of the young cluster IC~2391, and the peak's location in absolute magnitude suggests an age of 35 Myr. The location of this feature may prove a useful chronometer for young clusters.

Strong spectral features, such as the onset of methane absorption at the L/T transition, produce significant effects in the $K$-band luminosity function. Both these features and peak D should be detectable in nearby open clusters with the current generation of wide-field infrared imagers and SIRTF, allowing comparison of the model predictions with clusters spanning a much wider range in age.

P.R.A. acknowledges support by a grant made under the auspices of  the NASA/NSF NStars initiative, administered by JPL, Pasadena, CA. D.E.T. acknowledges support from NASA through a Space Telescope Science Institute grant to Gary Bernstein.  Thanks to Adam Burrows for making his models available for our use and for answering our questions about them.  P.R.A. thanks Kelle Cruz for her help in perfecting the plots in this paper.  The authors would also like to thank the referee for his/her valuable comments and quick response, as well as our editor, Paula Szkody, for her timeliness and excellent choice of referee.

\clearpage

\begin{deluxetable}{ccccccc}
\tablewidth{0pt}
\tablecaption{Cluster Information}
\tablehead{
\colhead{Cluster} & \colhead{MS turnoff age} & \colhead{Li-depletion age} & \colhead{Distance} & \colhead{Candidates} & \colhead{Ref.} \\
\colhead{Name} & \colhead{(Myr)} & \colhead{(Myr)} & \colhead{(pc)} & \colhead{(\#)} \\
}
\startdata
Upper Sco      & 5   & \nodata & 145 & 138 & (1) \\
IC 2391        & 35  & 53      & 155 & 132 & (2) \\
Pleiades       & 100 & 120     & 120 & 62  & (3) \\
\enddata
\tablerefs{(1) \citet{ard00}; (2) \citet{bn01}; (3) \citet{bouv}}
\end{deluxetable}

\clearpage

\begin{figure}
\rotatebox{-90}{
\epsscale{0.8}
\plotone{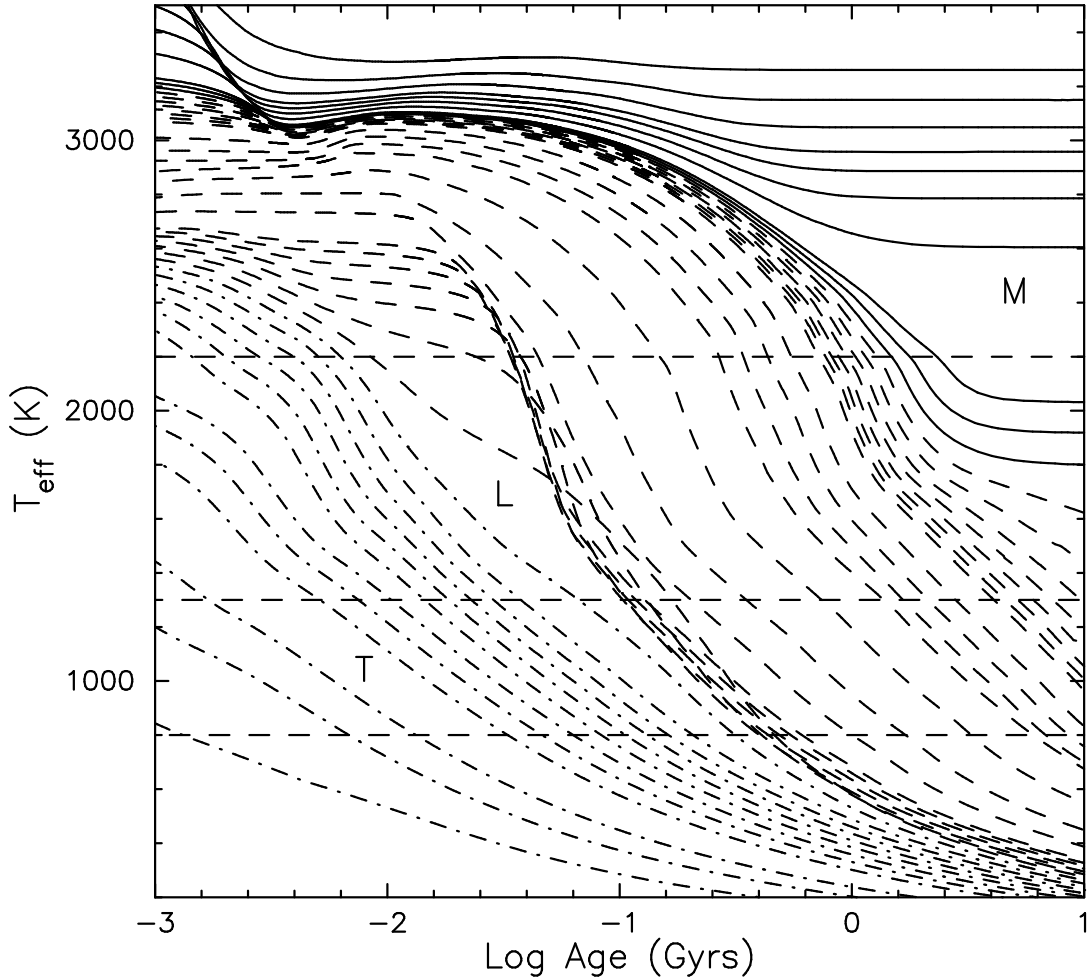}}
\figurenum{1}
\caption{\citet{bur} model evolutionary tracks as a function of effective temperature and age.  Solid curves show main sequence objects ($M > 0.075~M_{\odot}$); dashed curves show substellar objects massive enough to burn deuterium but not hydrogen ($13~M_J < M < 0.075~M_{\odot}$); and dot-dashed curves represent objects that never experience any significant thermonuclear reactions ($M < 13~M_J$).  Also marked are the ranges of MLT spectral types.  The contours correspond to the following object masses: 1, 2, 3, 5, 6, 7, 8, 9, 10, 11, 12, 13, 15, 16, 17, 18, 19~$M_J$, and 0.02, 0.025, 0.03, 0.035, 0.04, 0.05, 0.055, 0.06, 0.065, 0.07, 0.071, 0.072, 0.073, 0.074, 0.075, 0.076, 0.077, 0.078, 0.079, 0.08, 0.085, 0.09, 0.095, 0.1, 0.11, 0.125, 0.15~$M_{\odot}$ (1 $M_J = 1047 M_{\odot}$).}
\end{figure}

\clearpage

\begin{figure}
\centering
\rotatebox{-90}{
\epsscale{0.8}
\plotone{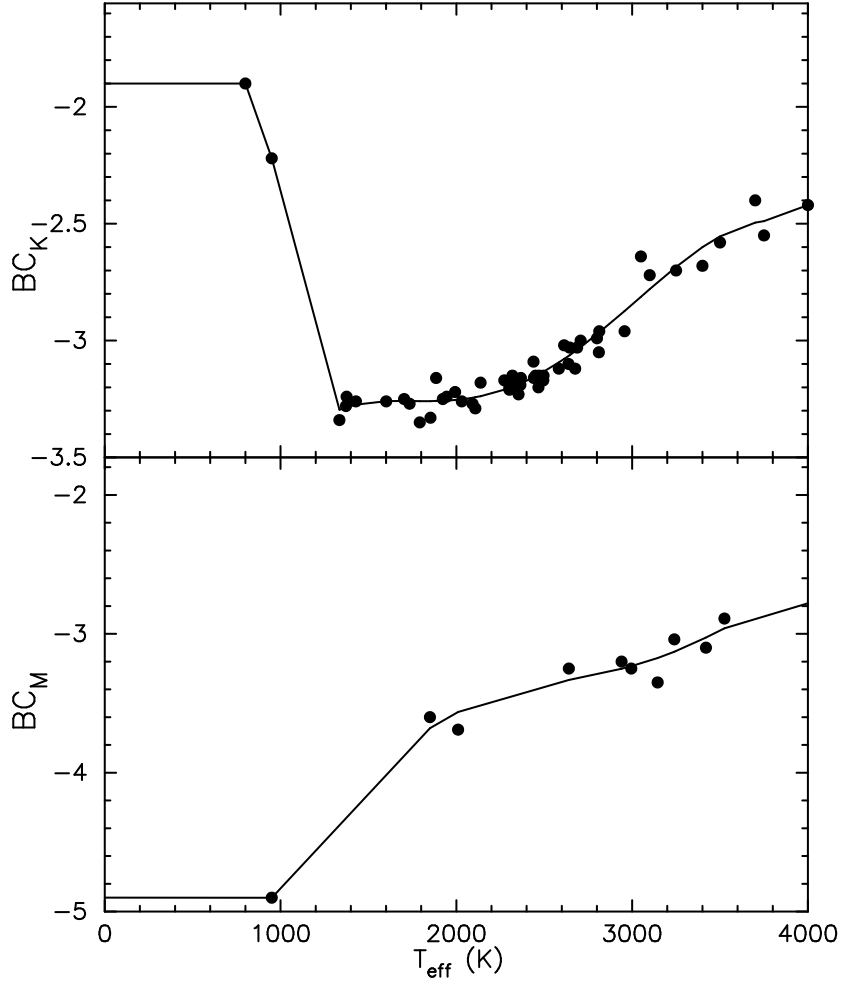}}
\figurenum{2}
\caption{Bolometric correction curves for $K$- and $M$-bands. The filled circles show individual measurements. The $K$-band data is based on the $J$-band corrections in \citet{dahn} and the $M$-band data is from \citet{rc}.  The curves are a seventh order polynomial fit to the data.  However, the T dwarfs (the coolest one $M$-band and two $K$-band points) are joined by straight line segments to the rest of the fit to preserve the sharpness of the L/T transition. The abrupt change in slope of the BC$_K$ data around 1300~K occurs at the L/T transition and is primarily due to methane absorption at temperatures cooler than 1300~K.  For temperatures cooler than T$\sim$900~K we assume the BCs are constant.}
\end{figure}

\clearpage

\begin{figure}
\centering
\epsscale{0.75}
\plotone{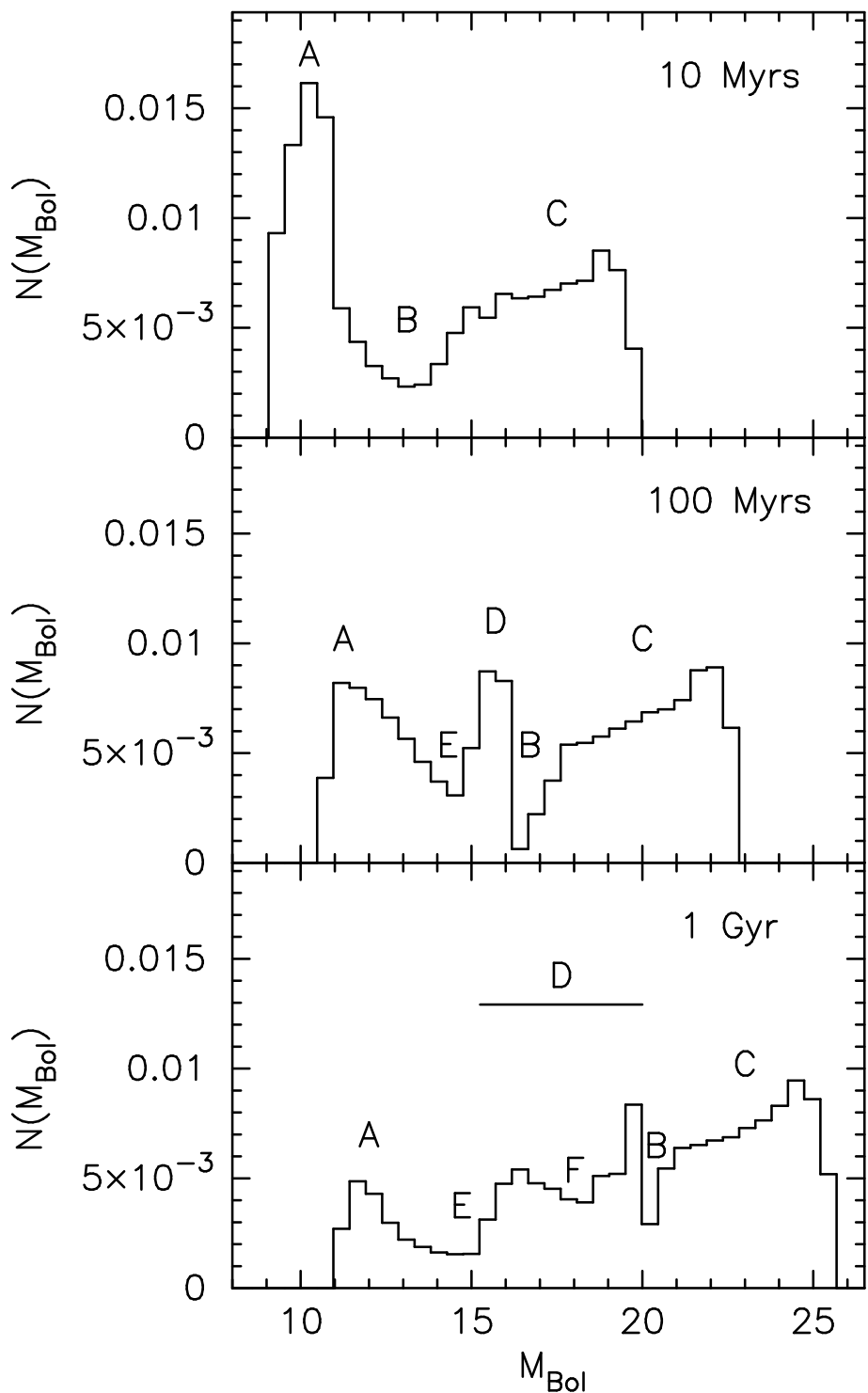}
\figurenum{3}
\caption{Modeled bolometric magnitude cluster luminosity function for three ages (10,~100,~and 1000~Myr) with $\alpha_2 = 1.0$.  Each feature is labeled as discussed in $\S$3.1.  Note that the sharp drop on the bright side of peak A is unphysical and is a result of the high mass edge of the Burrows models.  The normalization of the y-axis is dependent on the overall normalization of the distribution from which the luminosity function was created ($\S$2.1).}
\end{figure}

\clearpage

\begin{figure}
\centering
\rotatebox{-90}{
\plotone{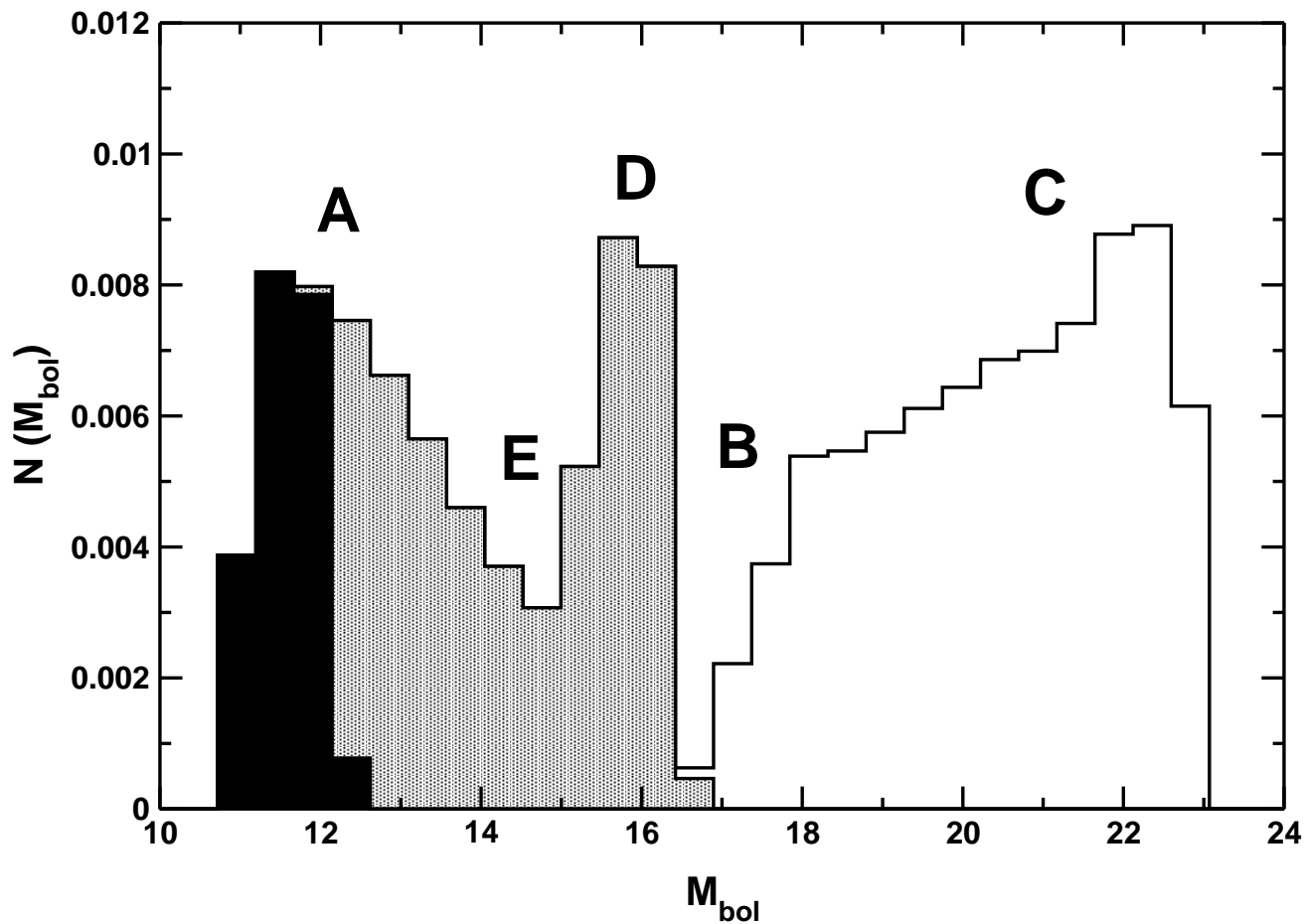}}
\figurenum{4}
\caption{The same as the middle panel of Figure 3 (100~Myr) for three different low-mass cutoffs.  The black shading corresponds to objects with $0.075~M_{\odot} < M \le 0.15~M_{\odot}$, the gray shading to $13~M_J < M \le 0.075~M_{\odot}$, and the unshaded to $M \le 13~M_J$.  This figure demonstrates that the intermediate peak, D, is populated solely by objects that burned deuterium.}
\end{figure}

\clearpage

\begin{figure}
\centering
\rotatebox{90}{
\epsscale{1}
\plottwo{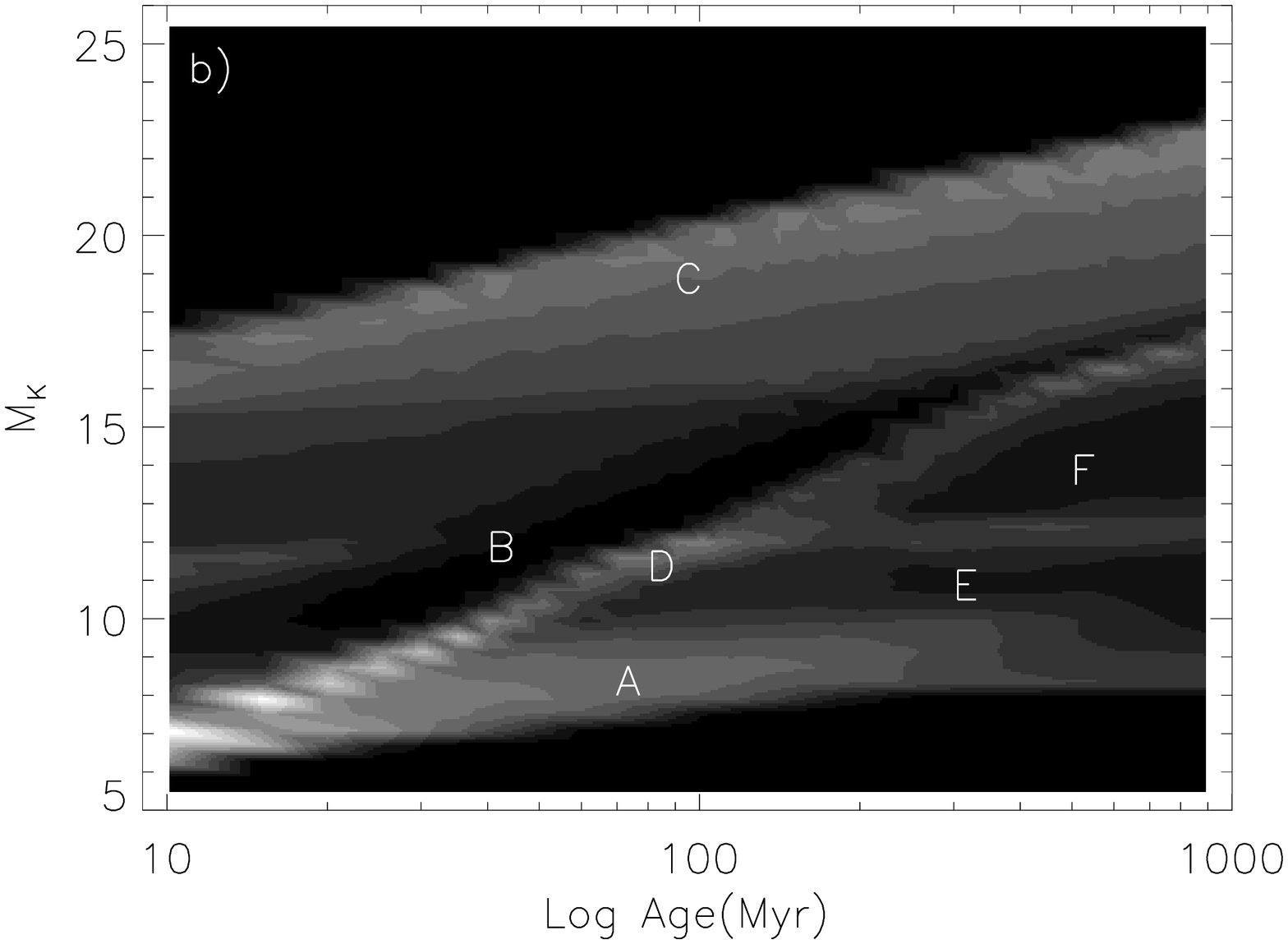}{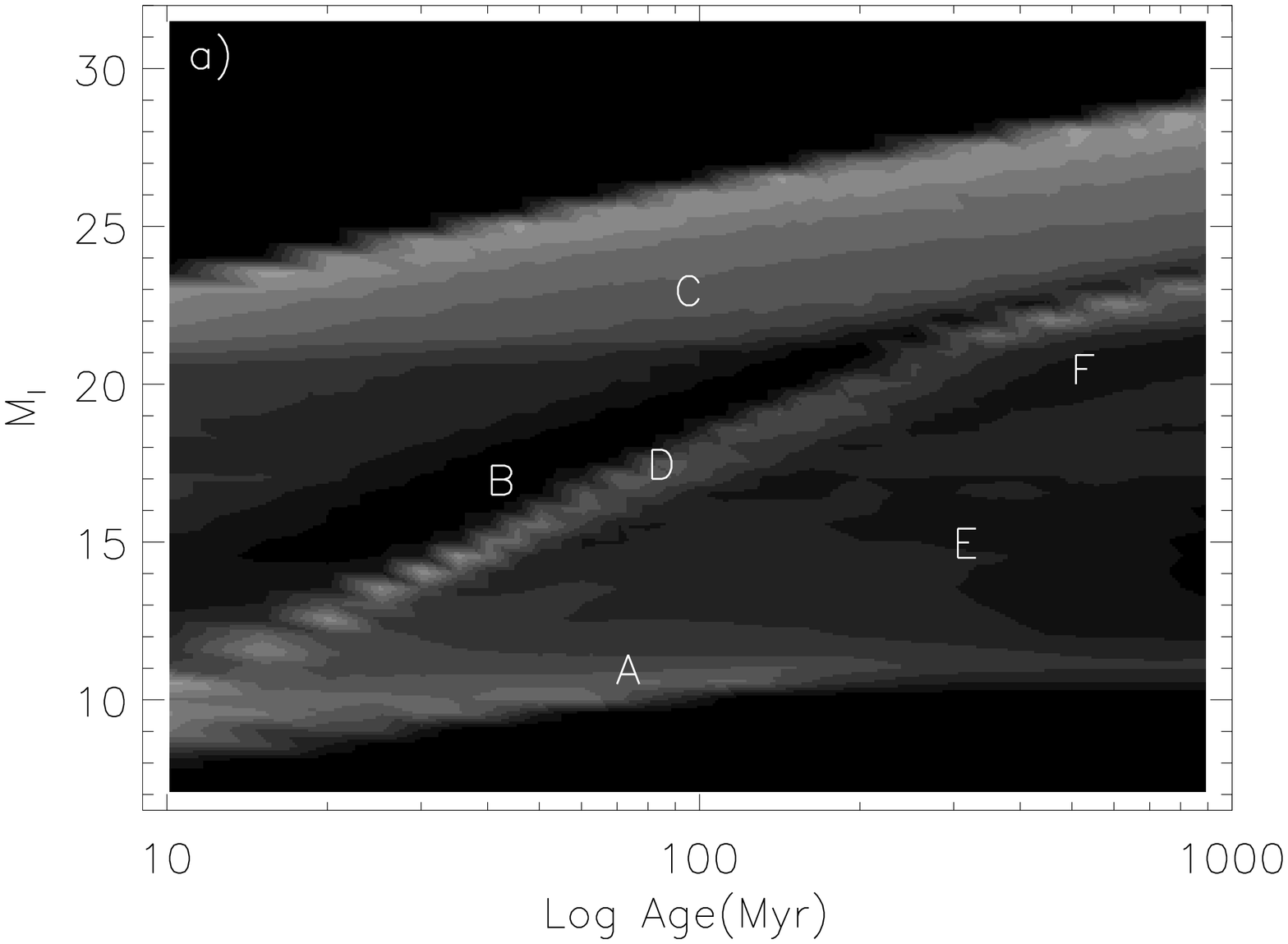}}
\figurenum{5}
\caption{Cluster luminosity function evolutions for $I$- (a) and $K$-bands (b). Light areas indicate relative high numbers, and dark areas indicate low numbers of objects.  The same feature labels used in Figure~3 are used here.  See the discussion in $\S3.2$ for details on each feature.  Each figure has the same grayscale normalization.}
\end{figure}

\clearpage

\begin{figure}
\centering
\rotatebox{-90}{
\epsscale{0.4}
\plotone{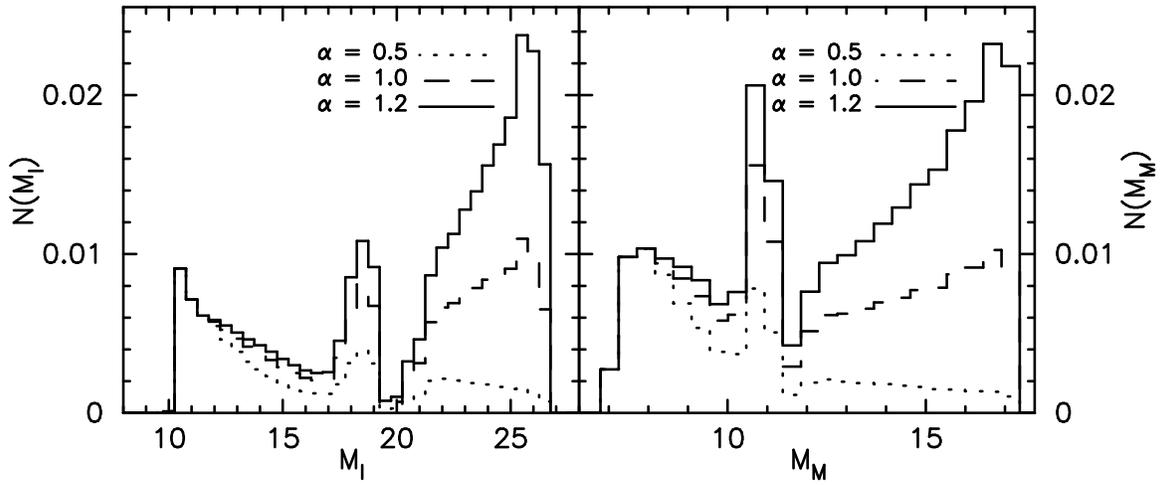}}
\figurenum{6}
\caption{$I$- and $M$-band luminosity functions at 100~Myr for three different power laws.  As expected, increasing $\alpha_2$ greatly increases the relative number of brown dwarfs compared to stars.  For $\alpha_2 = 0.5$ the substellar luminosity function slopes steeply down from the main sequence.}
\end{figure}

\clearpage

\begin{figure}
\centering
\epsscale{0.6}
\plotone{f7.ps}
\figurenum{7}
\caption{$I$-band data (triangles) for Upper Scorpius \citep{ard00}, IC~2391 \citep{bn01}, and the Pleiades \citep{bouv}, compared to $I$-band models for 5, 35/53, and 100/120 Myr old clusters, respectively, with $\alpha_{2} = 0.5$.  Absolute magnitudes are calculated based on the reported distances to each cluster.  We also show the Pleiades data from the photographic survey of Hambly, Hawkins, and Jameson 1993 (diamonds).  The models are normalized to the data points indicated by the solid circles as discussed in $\S$4.2.  Also, the first two to three model bins are not plotted as discussed in $\S$4.2.  We show two different age estimates for IC~2391, 35 Myr (solid) and 53 Myr (dotted), and the Pleiades, 120 Myr (solid) and 100 Myr (dotted).  Peak D appears to be detected in the IC~2391 data and distinguishes between the two age estimates.}
\end{figure}

\clearpage

\begin{figure}
\centering
\rotatebox{-90}{
\epsscale{1}
\plotone{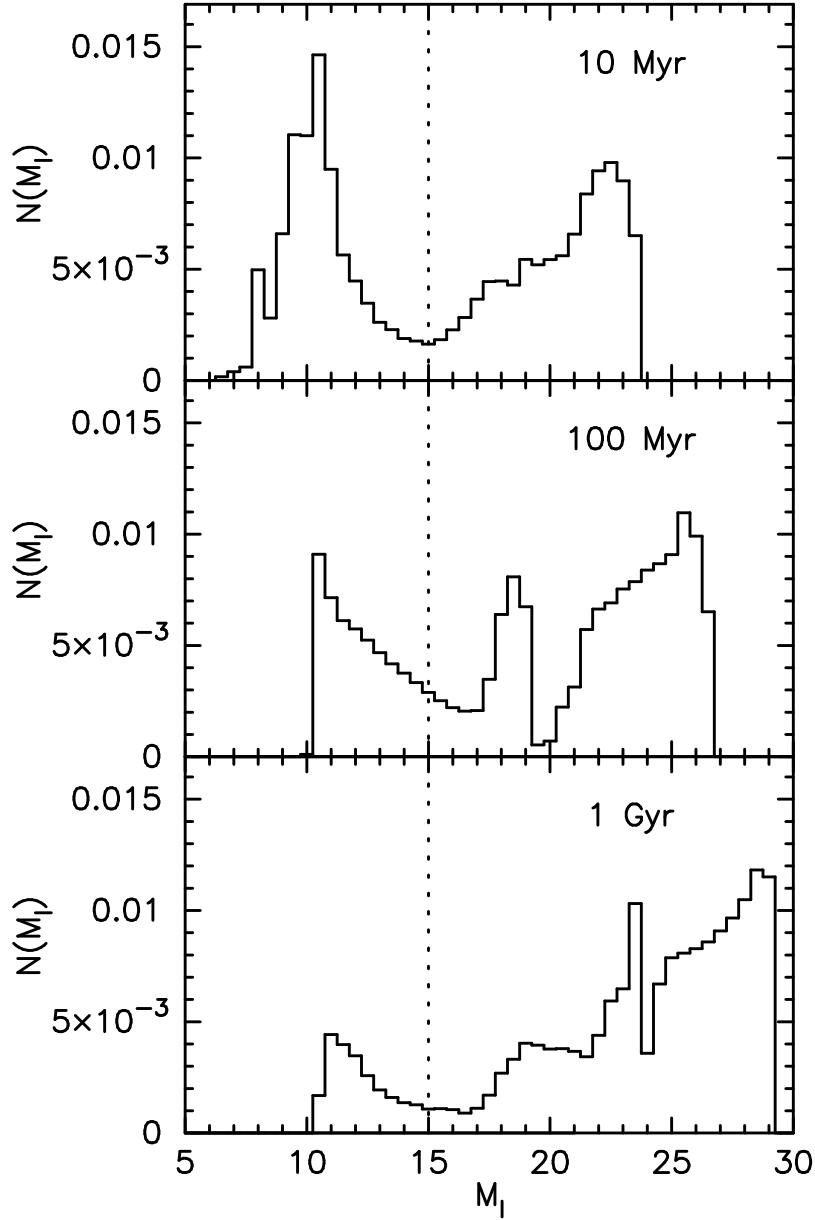}}
\figurenum{8}
\caption{Model $I$-band cluster luminosity functions for $\alpha_{2} = 1.0$ and ages 10, 100, and 1000 Myr with the $I$-band sensitivity limit from \citet{bouv} (dotted line).  This limit corresponds to 90\% completeness in 1200 seconds using the University of Hawaii 8K CCD mosiac at the CFHT for a cluster at 200 pc.  Note that this sensitivity limit lies in the middle of the stellar/substellar trough for each of the model clusters.  This implies that most ground-based $I$-band searches are unable to detect peak D except in relatively young nearby clusters.}
\end{figure}

\clearpage

\begin{figure}
\centering
\rotatebox{-90}{
\epsscale{1}
\plotone{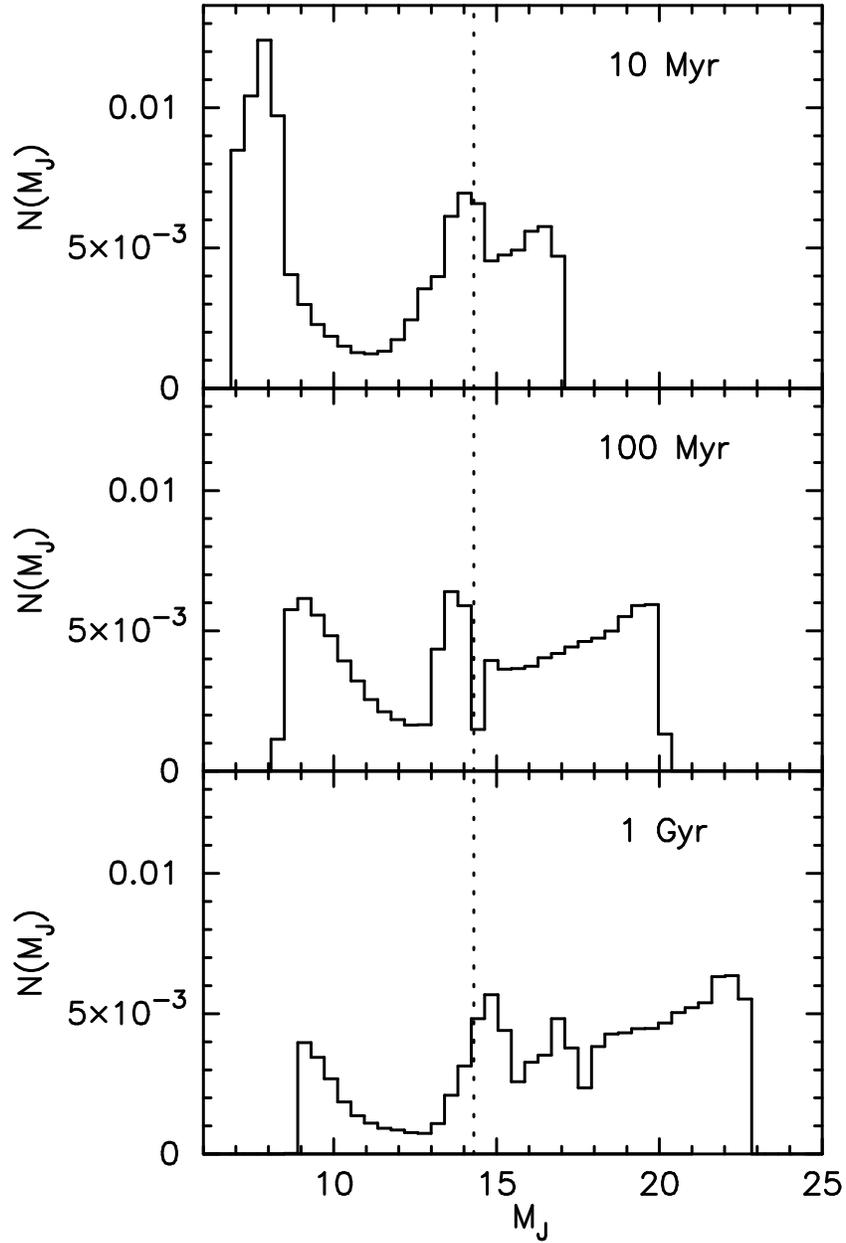}}
\figurenum{9}
\caption{Similar to Figure 8, but for $J$-band.  The sensitivity limit is for a cluster at 200~pc using Flamingos at Gemini South which can reach $m_J = 20.3$ in 300~sec for a 5$\sigma$ point source.}
\end{figure}

\clearpage

\begin{figure}
\centering
\rotatebox{-90}{
\epsscale{1}
\plotone{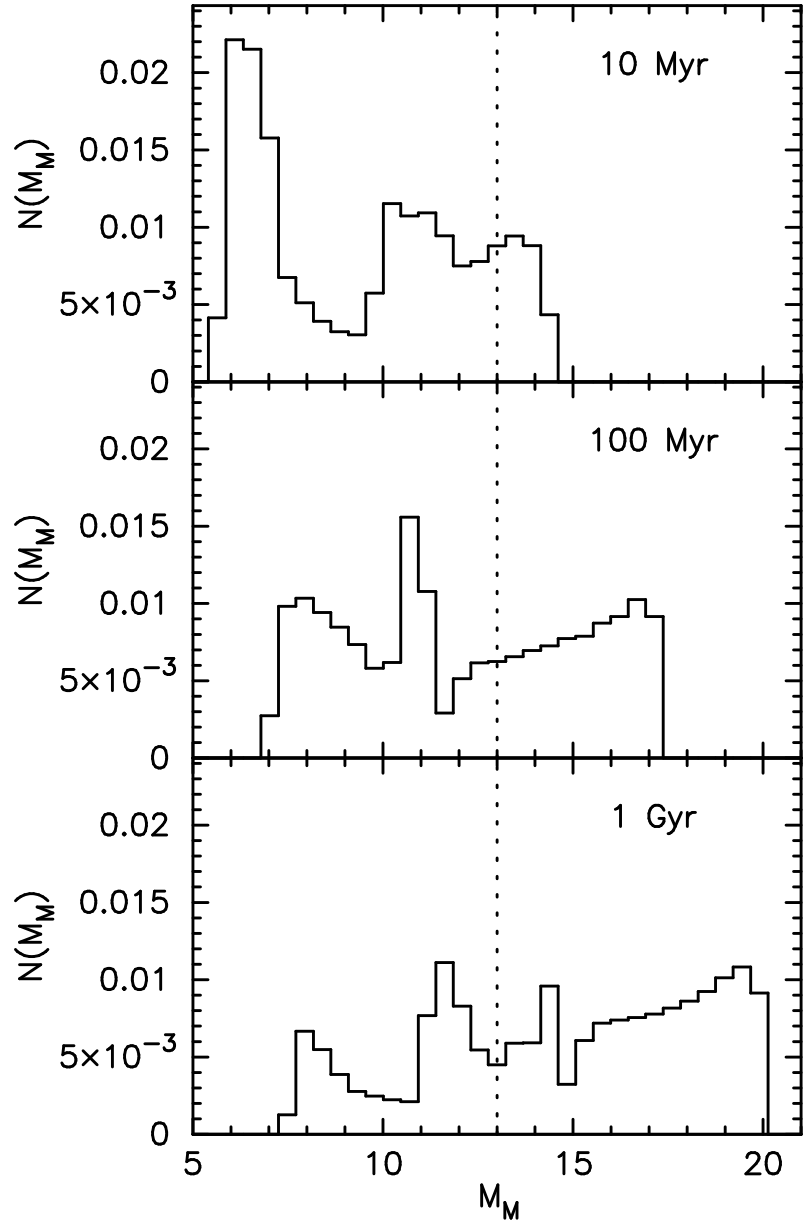}}
\figurenum{10}
\caption{Similar to Figure 8, but for $M$-band.  The sensitivity limit corresponds to a 5$\sigma$ point source detection in 200 seconds with IRAC Channel 2 on SIRTF for a cluster at 200~pc.}
\end{figure}

\end{document}